\documentclass[RNAAS]{aastex62}

\accepted{January 2019, Volume 3, Issue 1}

\begin{document}

%\title{Statistically low amplitude molecular absorption features in exoplanet atmospheres}
\title{Exoplanet Atmosphere Forecast: Observers Should Expect Spectroscopic Transmission Features to be Muted to 33\%}

%% Note that the corresponding author command and emails has to come
%% before everything else. Also place all the emails in the \email
%% command instead of using multiple \email calls.
\correspondingauthor{H.R. Wakeford}
\email{hwakeford@stsci.edu}

\author[0000-0003-4328-3867]{H.R. Wakeford}
\altaffiliation{Giacconi Fellow}
% \affiliation{Space Telescope Science Institute, Baltimore, MD 21218, USA}
\author[0000-0001-6352-9735]{T.J. Wilson}
% \affiliation{Space Telescope Science Institute, Baltimore, MD 21218, USA}
\author[0000-0002-7352-7941]{K.B. Stevenson}
\affiliation{Space Telescope Science Institute, 3700 San Martin Drive, Baltimore, MD 21218, USA}
\author[0000-0002-8507-1304]{N.K. Lewis}
\affiliation{Department of Astronomy and Carl Sagan Institute, Cornell University, 122 Sciences Drive, Ithaca, NY 14853, USA}

%% Note that RNAAS manuscripts DO NOT have abstracts.
\keywords{planets and satellites: atmospheres, methods: statistical}

%% Start the main body of the article. If no sections in the 
%% research note leave the \section call blank to make the title.
\section{} 
To ensure robust constraints are placed on exoplanet atmospheric transmission spectra, future observations need to obtain high signal-to-noise ratio (SNR) measurements assuming smaller amplitude molecular signatures than those of clear solar metallicity atmospheres. Analyzing 37 exoplanet transmission spectra we find clear solar molecular features are measured in $<$7\% of cases. Clear solar metallicity atmospheres, free from cloud opacities, represent the highest SNR scenario to measure molecular absorption features in exoplanet transmission spectra and are often the maximum assumed atmospheric signal. Some of the most prominent absorption features expected in close-in giant exoplanet spectra are those of water. The distinct absorption feature of H$_2$O at 1.4$\mu$m has been observed in multiple exoplanet transmission spectra (e.g., \citealt{sing2016,wakeford2018}) using the Hubble Space Telescopes (HST) Wide Field Camera 3 (WFC3) G141 grism. However, many show little or no water absorption (e.g., \citealt{sing2016,kreidberg2014a}). In this study we determine the statistical significance on the muting of molecular absorption features as a population. The purpose of this study is to perform a global analysis, not to present detailed analyses of each planet. 

\begin{figure*}[t]
\begin{center}
\includegraphics[width=0.87\textwidth]{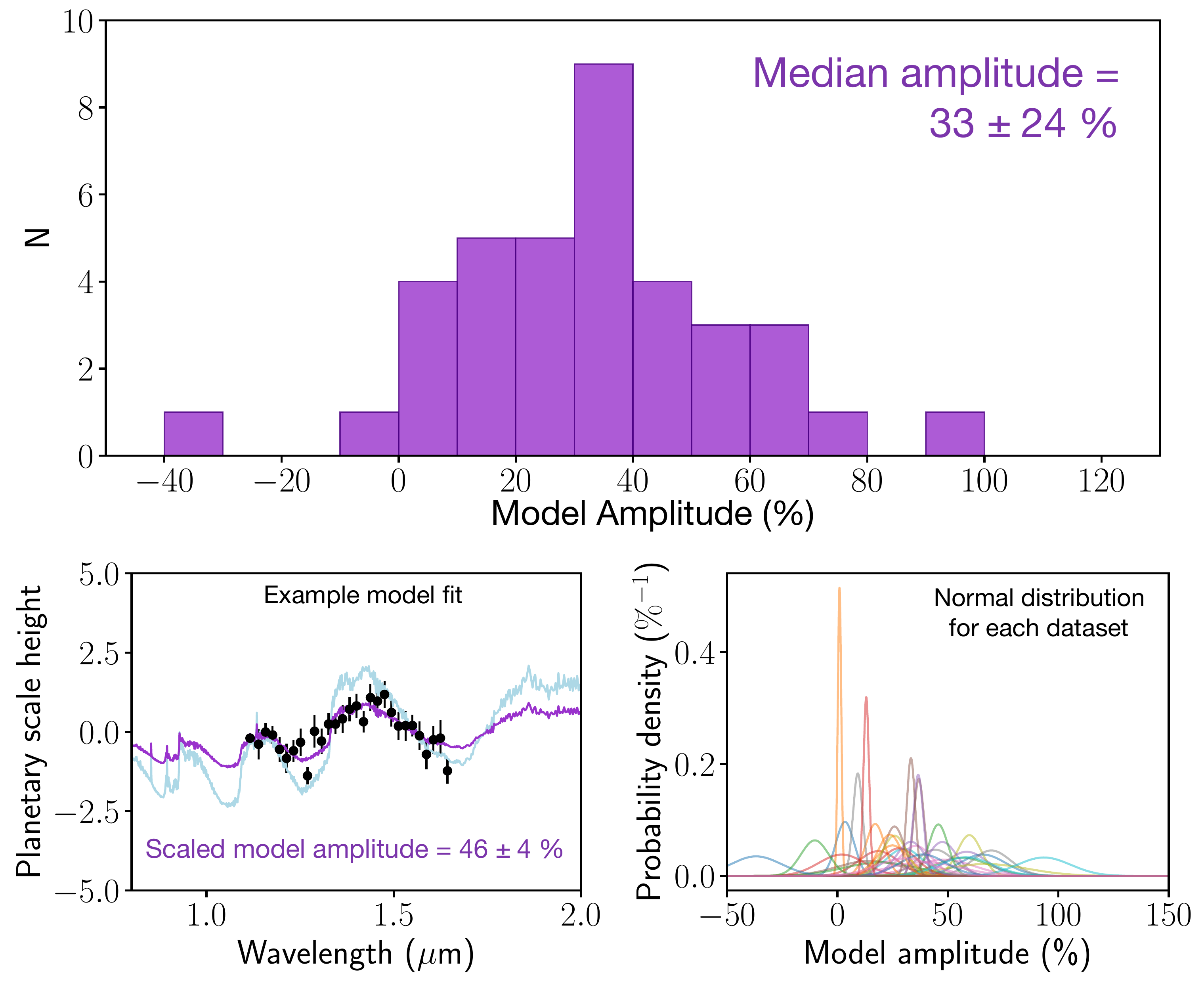}
\caption{Top: histogram of amplitudes of molecular absorption features in 37 exoplanet transmission spectra with HST WFC3 G141 grism. The median amplitude of absorption features is found to be muted to 33$\pm$24\% of expected clear solar models. Bottom left: example of the scaled model (purple) fit to data (\citealt{wakeford2018}, black points) to measure the relative amplitude of the feature compared to a clear solar model (blue). Bottom right: probability density for model fits to each exoplanet sampled.}
\label{fig:plots}
\end{center}
\end{figure*}

We find the amplitude of molecular absorption features, 1.1--1.7\,$\mu$m, is muted to 33\,$\pm$\,24\,\% of the expected clear solar atmospheric models (Fig. \ref{fig:plots}, top). To obtain this, we compiled measurements of 37 exoplanets observed with HST WFC3 G141 (across 16 programs) in transmission. The data are all publicly available, but for unpublished datasets we use methods outlined in \citet{wakeford2016} and \citet{stevenson2014a} to measure the transmission spectrum. 

We fit each transmission spectrum with a 1D isothermal model from the generic ATMO grid \citep{Goyal2019}, where the model has solar metallicity and C/O ratio with no scattering or uniform opacity sources, to represent the best case scenario for atmospheric absorption. Each model is scaled to the individual planetary parameters based on the stellar radius, planetary radius, equilibrium temperature ($T_{eq}$), and planetary gravity ($g_p$)\footnote{obtained from TEPCat http://www.astro.keele.ac.uk/jkt/tepcat/tepcat.html}. We use the $T_{eq}$ and $g_p$ to select the most appropriate model from the grid, and rescale the model to the specific planetary parameters. To fit the model to the data we use a least-squares minimizer with the model defined as 
$S1 = (S0 \times p_0) + p_1$, 
where $S1$ is the scaled model, $S0$ the clear solar model, $p_0$ the model amplitude scale factor , and $p_1$ a baseline offset; we also calculate the uncertainty $\sigma_\mathrm{p_0}$. As expected, there is a correlation between the uncertainty on the model amplitude and the mean uncertainty on the transmission spectral data. Conversely, we find no correlation between the SNR of the atmospheric transmission and the uncertainty on the model amplitude. An example model fit is shown in Fig. \ref{fig:plots}, bottom left. 

To determine the collective distribution of model amplitudes across all datasets we use $p_0$ and $\sigma_\mathrm{p_0}$, fitting as a Gaussian distribution from -60\,--\,150\% (Fig. \ref{fig:plots}, bottom right). From each normal distribution we randomly sample 5000 points to determine the global 16$^\mathrm{th}$, 50$^\mathrm{th}$, and 84$^\mathrm{th}$ percentile bounds. We find the median and 1$\sigma$ bounds on the amplitude of molecular transmission features are 33\,$\pm$\,24\,\%. The model percentage amplitude forms a one-to-one correlation with atmospheric scale heights (H) with the median distribution equivalent to 0.89$\pm$0.77\,H, encompassing the more conservative value of 1.4\,H presented by \citet{Fu2017}. Statistically, 30\% of the time the amplitude of the observed feature is below 20\% or 0.5H, and an approximate clear solar metallicity atmosphere, $\ge$70\% or $\ge$2H, is measured only 7\% of the time. 

In summary, we find a median absorption feature amplitude that is one-third the strength derived assuming a clear solar metallicity atmosphere. Our analysis demonstrates that assumptions of clear solar metallicity atmospheres for exoplanets will overestimate the achievable SNR and potentially underestimate observational time requirements $\sim$93\% of the time. To ensure robust constraints on exoplanet atmospheric properties, future observations should account for the likelihood that atmospheric absorption features, particularly in transmission, may be muted due to a broad range of atmospheric properties and processes.

\acknowledgments
% Acknowledge people, facilities, and software here but remember that this counts
% against your 1000 word limit.
Thanks to J. Valenti, D. Deming, D.K. Sing, and M. Lopez-Morales for useful comments. 
This work is based on observations made with the NASA/ESA HST that were obtained at the STScI, operated by AURA. H.R.Wakeford acknowledges support from the AURA/STScI Giacconi Fellowship.
% from the following programs: GO-11622, GO-12181,GO-12449, GO-12473, GO-12881,GO-12956,GO-13021, GO-13431,GO-13467,GO-13501,GO-14260, GO-14468, GO-14619, GO-14642, GO-14767, GO-14915; 

% \bibliographystyle{aasjournal}
% \bibliography{Wakeford_references}

\end{document}